\documentclass[aps,prd,preprint,showpacs,nofootinbib,superscriptaddress,tightenlines]{revtex4}
\usepackage{amsmath,amssymb,amsbsy,color,psfrag}  

\newcommand{\be}{\begin{equation}}
\newcommand{\ee}{\end{equation}}
\newcommand{\bea}{\begin{eqnarray}}
\newcommand{\eea}{\end{eqnarray}}
\newcommand{\bi}{\begin{itemize}}
\newcommand{\ei}{\end{itemize}}
\newcommand{\non}{\nonumber}
\newcommand{\tr}{{\rm tr}}

\begin{document}

%\draft
\preprint{UTHEP-638}
\preprint{UTCCS-P-65}
\preprint{KANAZAWA-11-19}

\title{
On the phase of quark determinant
in lattice QCD with finite chemical potential
}

% Takeda
\author{Shinji Takeda}
\affiliation{
Institute of Physics, Kanazawa University, Kanazawa 920-1192, Japan
}
\affiliation{
RIKEN Advanced Institute for Computational Science,
Kobe, Hyogo 650-0047, Japan
}
% Kuramashi
\author{Yoshinobu Kuramashi}
\affiliation{
RIKEN Advanced Institute for Computational Science,
Kobe, Hyogo 650-0047, Japan
}
\affiliation{
Graduate School of Pure and Applied Sciences,
University of Tsukuba,
Tsukuba, Ibaraki 305-8571, Japan
}
\affiliation{
Center for Computational Sciences,
University of Tsukuba,
Tsukuba, Ibaraki 305-8577, Japan
}
% Ukawa
\author{Akira Ukawa}
\affiliation{
Center for Computational Sciences,
University of Tsukuba,
Tsukuba, Ibaraki 305-8577, Japan
}
\date{\today}
%
%===========
\begin{abstract}
%===========
%
We investigate the phase of the quark determinant with finite chemical potential
in lattice QCD using both analytic and numerical methods.
Applying the winding number expansion and the hopping parameter expansion to the logarithm of the determinant,
we show that the absolute value of the phase has
an upper bound that grows with the spatial volume but decreases exponentially
with an increase in the temporal extent of the lattice.
This analytic but approximate result is confirmed
with a numerical study in four-flavor QCD 
in which the phase is calculated exactly.
Since the phase is well controlled on lattices with larger time extents,
we try the phase reweighting method in a region beyond $\mu/T=1$
where the Taylor expansion method cannot be applied.
Working in four-flavor QCD, we find a first-order like behavior  on a $6^3\times 4$ lattice at $\mu /T\approx 0.8$ 
which was previously observed by Kentucky group with the canonical method.
We also show that the winding number expansion has a nice convergence property
beyond $\mu/T=1$.     
We expect that this expansion
is useful to study the high density region of the QCD phase diagram at low temperatures. 
% therefore we propose a strategy which exploits its nice property.

\end{abstract}

\pacs{11.10.Wx, 12.38.Gc}
\maketitle

%1
%=================
\section{Introduction}
%=================
%

Establishing the QCD phase diagram spanned by the temperature $T$ and the quark chemical potential $\mu$
in a quantitative way
is an important task of lattice QCD.
See Refs.~\cite{Gupta:2011ma,deForcrand:2010ys}
for recent progress.
The Monte Carlo simulation technique, which has been
successfully applied to the finite temperature phase transition studies in lattice QCD,
cannot be directly applied to the finite density case 
due to the complexity of the quark determinant $\det D=|\det D|e^{i\theta}$ for $\mu\ne 0$. 
A simple way out of the problem is the reweighting method
which incorporates the absolute value of the determinant $|\det D|$ into the integral measure.  This bipass, however,
suffers from the sign problem with increasing $\mu$ 
caused by increasingly larger gauge fluctuations of the reweighting factor $e^{i\theta}$.
Understanding the property of the phase is crucial to control the sign problem.

The average of the phase factor $\langle e^{i\theta}\rangle$
was investigated in the framework of chiral perturbation theory in Refs.~\cite{
Splittorff:2006fu,Splittorff:2007ck,Splittorff:2007zh}
and by random matrix model in Ref.~\cite{Han:2008xj} thus far.
The phase itself $\theta$ has also been investigated 
by Taylor expansion in lattice QCD \cite{Allton:2002zi,Sasai:2003py,Ejiri:2004yw,Ejiri:2005ts}.
In this article we report on our attempt to understand the property of the phase directly in lattice QCD.
Near the phase transition, low energy effective theories may not be reliable.  To survey such a region, it is imperative to work within the first principle framework of QCD itself. 

The Taylor expansion method works directly with QCD.  On the $T$-$\mu$ plane of QCD, however, we expect the expansion to be reliable only in the region $\mu/T<1$.
Therefore, alternative approximation methods which works for 
wider region in the parameter $\mu/T$ is a welcome step 
to understand the phase diagram.
In this paper, we show that
the winding number expansion~\cite{Danzer:2008xs}, whose convergence is better for larger temporal size of lattice,
can be a possible candidate of the approximation.

This expansion, combined with the hopping parameter expansion, allows us an analytical investigation of 
the lattice parameter dependence of the phase
for the determinant of the Wilson operator.
The result, which is obtained at the leading order of the hopping parameter expansion, tells us that the magnitude of the phase decreases
for larger temporal size of the lattice with other parameters held fixed.
A numerical test in $N_{\rm f}=4$ QCD with an exact calculation of the phase has confirmed that this property holds true beyond the leading order of the hopping parameter expansion.

These analyses drive us to try a further numerical study with the reweighting method since we can avoid the sign problem by increasing the temporal size of the lattice.
As a first testing ground, we carry out $N_{\rm f}=4$ QCD simulation and observe a first-order like behavior
which was previously reported in Ref.~\cite{Li:2010qf}.

The rest of the paper is organized as follows.
In Sec. \ref{sec:phase},
we discuss 
the convergence of the winding number expansion, and
the behavior of the phase as a function of lattice parameters.
Simulation results in $N_{\rm f}=4$ QCD are shown in Sec. \ref{sec:RW}.
Finally, we present concluding remarks.
  
Throughout this paper we consider a 4-dimensional Euclidean lattice of a size $N_{\rm L}^3\times N_{\rm T}$
with the periodic (anti-periodic) boundary conditions in the time direction for boson (fermion) fields.
A summary of the notations and definitions relevant for the winding number expansion and a reduction process of the quark determinant is given in appendix~\ref{sec:wilsonoperator}.

%2
%=================================
\section{Phase of quark determinant}
\label{sec:phase}
%=================================
%
%After examining the convergence properties of the winding number expansion,
%we derive an upper bound of the absolute value of the phase
%on the basis of the hopping parameter expansion.

\subsection{Winding number expansion}
\label{subsec:WNE}
The starting point of our discussion is the winding number expansion of the logarithm of the quark determinant for Wilson-type fermion action~\cite{Danzer:2008xs}.  A brief review of the method is provided in appendix~\ref{sec:wilsonoperator} for the case of the unimproved Wilson fermion action.  
Inclusion of the clover term is straightforward, however, and the 
simulation results presented later in this article are obtained for the clover-improved fermion action.

For the one-flavor case, the winding number expansion takes the form, 
\bea
\det [D(\kappa,\mu)]
&=&
A_0(\kappa)\exp
\left[
-\sum_{q\in\mathbb{Z}}e^{q\mu/T}V^{(q)}(\kappa)
\right],
\label{eqn:WNE}
\eea
where the dependence on the hopping parameter $\kappa=1/(8+2am_0)$ and the quark chemical potential $\mu$ 
is given explicitly. 
In the original paper \cite{Danzer:2008xs}, the authors used $T^{(q)}$ instead of $V^{(q)}$.  $T=1/(aN_{\rm T})$ is the temperature.
As explained in appendix~\ref{sec:wilsonoperator}, 
the above expression is derived by casting the Wilson-Dirac operator into blocks connecting time slices, and reducing the quark determinant in terms of these blocks.
%This shows that the determinant is factorized into two parts:
%The local and global quark dynamics for the time direction are contained in
%the first and second factor respectively.

The first factor $A_0$ defined in eq.~(\ref{eqn:A0})
is composed from mainly the block diagonal parts of the Wilson operator in a time blocked form.  Hence it depends on $\kappa$ but is independent of $\mu$.  
This factor can be shown to be real, but is not guaranteed to be positive.
On the other hand, $V^{(q)}(\kappa)$ ($q\neq0$) in the second factor, which are defined in eq.~(\ref{eqn:Tq}), contain global information in the time direction, being composed from the off-diagonal parts of the Wilson operator.
Namely, $V^{(q)}(\kappa)$ is a sum of quark loops winding around the time direction $q$ times, and so appears with the $q$-th power of fugacity  $e^{\mu/T}$.  
It has the following properties:
\bea
V^{(0)}&\in&\mathbb{R},
\label{eqn:Tqast0}
\\
V^{(q)}&\in&\mathbb{C},
\hspace{3mm}
V^{(-q)}=V^{(q)^\ast}\mbox{ for } q\neq 0.
\label{eqn:Tqast}
\eea
The details are given in appendix~\ref{sec:wilsonoperator}.
%The global part $V^{(q)}$ with $q\neq0$
%is closely related with the fugacity $e^{\mu/T}$.

Let us express the phase of the determinant in terms of the winding number expansion.
To this end, using eqs.~(\ref{eqn:Tqast0}) and (\ref{eqn:Tqast}),  we rewrite,  
\be
\det [D(\mu)]
=
A_0e^{-V^{(0)}}
\exp
\left[
-\sum_{q=1}^{\infty}
\left(
2\cosh(q\mu/T){\rm Re}[V^{(q)}]
+
i2\sinh(q\mu/T){\rm Im}[V^{(q)}]
\right)
\right],
\label{eqn:onedet}
\ee
and defining the phase through
\be
\det [D(\mu)]
=
|\det [D(\mu)]|e^{i\theta(\mu)},
\ee
we read
\bea
|\det [D(\mu)]|
&=&
|A_0|e^{-V^{(0)}}
\exp
\left[
-\sum_{q=1}^{\infty}
2\cosh(q\mu/T){\rm Re}[V^{(q)}]
\right],
\\
\theta(\mu)
&=&
\arg(A_0)
-\sum_{q=1}^{\infty}
2\sinh(q\mu/T){\rm Im}[V^{(q)}].
\eea

Since $A_0$ is a real number, the phase contribution $\arg(A_0)$ takes the value $0$ or $\pi$.
For even number of degenerate flavors, it is zero because of $\arg(A_0^2)=0$.
Even for odd number of flavors, 
we expect that $\arg(A_0)=0$ holds if the corresponding quark mass is heavy enough,
since we know that the Wilson fermion determinant with $\mu=0$ is effectively positive in the strange mass region.
Therefore in the following we ignore
%\footnote{In the case of an odd number of flavor, one has to check the sign of $A_0$ in an actual simulation.
%In this paper, we deal with four flavor case thus we do not have to do it.}
the phase contribution from the local dynamics
$\arg(A_0)$ and exclusively consider the essential part of the phase
which depends on $\mu$ vanishing at $\mu=0$,
\be
\theta(\mu)
=
-\sum_{q=1}^{\infty}
2\sinh(q\mu/T){\rm Im}[V^{(q)}].
\label{eqn:thetaTq}
\ee
%This expression shows that the gauge fluctuation of the imaginary part of $V^{(q)}$,
%which is something like a fussy Polyakov loop
%winding around the time direction $q$ times, is magnified
%by the exponential factor $\sinh(q\mu/T)$
%and then the phase may be affected by large fluctuation.

For later use, we write down the phase for the general $N_{\rm f}$-flavor case.
From the full determinant,
\be
\prod_f \det [D(\kappa_f,\mu_f)]
=
\left|\prod_f\det [D(\kappa_f,\mu_f)]\right|\exp\left[i\sum_f\theta(\kappa_f,\mu_f)\right],
\ee
we read the total phase
\be
\theta_{\rm total}
=
\sum_f\theta(\kappa_f,\mu_f)
=
-\sum_f\sum_{q=1}^{\infty}2\sinh(q\mu_f/T){\rm Im}[V^{(q)}(\kappa_f)].
\ee
%The fluctuation of the total phase tends to be large for more flavors,
%since the total phase is a sum of each phase.

\subsection{Convergence of the winding number expansion}
%In this subsection, before discussing
%a property of the phase,
%we consider the convergence of the winding number expansion.

%Some readers may consider that the winding number expansion of
%the determinant is just a formal expansion in terms of a topological number
%and it does not converge in a practical sense.
%In the following, however, we show that this has good convergence property
%in a paticular parameter region.

%In the expression of eq.~(\ref{eqn:thetaTq}), it is not actually clear which is an expansion parameter.
In order to identify an effective expansion parameter in the winding number expansion eq.~(\ref{eqn:thetaTq}), we adopt two simplifications.
The first factor behaves as 
\be
2\sinh(q\mu/T)\sim\exp(q\mu/T),
\label{eqn:sinhscaling}
\ee
for large $q$.
For the second term, one expects 
\be
V^{(q)}
\propto
(2\kappa)^{qN_{\rm T}}
\label{eqn:Tqq}
\ee
for small $\kappa$ and large $q$, since the quark loops 
in $V^{(q)}$ have lengths of at least $qN_{\rm T}$ and each hop along the loop comes together with a factor of $2\kappa$.
Writing the proportionality constant in eq.~(\ref{eqn:Tqq}) as $c_q$, the phase is estimated as 
\be
\theta
\rightarrow
-\sum_{q=1}^{\infty}
e^{(q\mu/T)}
\cdot
c_q(2\kappa)^{qN_{\rm T}}
=
-\sum_{q=1}^{\infty}
c_q
\{e^{\mu/T+N_{\rm T}\ln(2\kappa)}\}^q.
\ee
Assuming that $c_q$ has a mild $q$-dependence,
this expansion converges when the ``effective"
expansion parameter $e^{\mu/T+N_{\rm T}\ln(2\kappa)}<1$ is small.
%namely $\mu/T+N_{\rm T}\ln(2\kappa)<0$.
Therefore, the convergence region is given by
\be
\mu/T<-N_{\rm T}\ln(2\kappa).
%\hspace{10mm}
%\mbox{ for }
%\hspace{10mm}
%\kappa\sim0.17 (\ln(2\kappa)\sim -1).
\ee
Since actual simulations are performed at $\kappa>0.125$ ($\ln(2\kappa)<-1.386\cdots$),
the convergence region is much wider than that of Taylor expansion, $\mu/T<1$.
This analysis also indicates that the convergence becomes better for larger $N_{\rm T}$
for $a\mu$ fixed.

\subsection{An upper bound for the absolute value of the phase}
\label{sec:upperbound}
The origin of the sign problem is a large fluctuation of the phase.
Understanding the properties of the phase in more detail may open a new insight on either solving or avoiding the sign problem.
Here we derive an analytical upper bound for the leading term
of the phase in the hopping parameter expansion.
While this is just an approximated upper bound,  it reveals interesting features of the phase, which we later confirm by numerical simulations. 
%so the reader may consider that the meaning of the bound is rather limited.
%However, we believe that it is still useful to obtain a rough behavior of the phase.
%Furthermore such behavior is observed in the actual simulation results, that is
%without the hopping parameter expansion.

To derive the upper bound,
we use the inequality $|x+y|\le |x|+|y|$ for $x,y\in\mathbb{C}$ repeatedly to find that 
%then the (exact) upper bound of
%the absolute value of the phase is given by
\be
|\theta|
\le
\sum_{q=1}^{\infty}
2\sinh(q\mu/T)
|{\rm Im}[V^{(q)}]|
\le
\sum_{q=1}^{\infty}
2\sinh(q\mu/T)
|V^{(q)}|.
\label{eqn:ubound_theta}
\ee
To obtain an explicit lattice parameter dependence, let us apply some approximations.
First we truncate the winding number expansion at $q=1$.
In the previous section,
this truncation has been shown to be valid for small $\kappa$ and large $N_{\rm T}$.
After this truncation, the upper bound becomes
\be
\sum_{q=1}^{\infty}
2\sinh(q\mu/T)
|V^{(q)}|
\longrightarrow
2\sinh(\mu/T)
|V^{(1)}|.
\ee
We then use the explicit expressions for $V^{(1)}$ given in eqs.~(\ref{eqn:Tq}) and (\ref{eqn:Tqn}), and obtain 
\bea
|V^{(1)}|
&=&
|{\rm tr}
(H_+)|
+O((2\kappa)^{2N_{\rm T}})
%\hspace{5mm}
%\because \mbox{eq.~}({\rm\ref{eqn:Tq}})
\non\\
&=&
(2\kappa)^{N_{\rm T}}
\cdot
{\rm tr}_{\rm spin}
[(P^{(4)})^{N_{\rm T}}]
\cdot
\left|
\sum_{\bf x}
{\rm tr}_{\rm color}
\left[
\prod_{x_4=1}^{N_{\rm T}}U_4({\bf x},x_4)
\right]
\right|
+O(N_{\rm T}(2\kappa)^{N_{\rm T}+2})
\non\\
&\le&
(2\kappa)^{N_{\rm T}}
\cdot
2
\cdot
3N_{\rm L}^3,
\label{eqn:upperT}
\eea
%Note that in the last step we have neglected $O(N_{\rm T}(2\kappa)^{N_{\rm T}+2})$
%(this approximation is valid for small $\kappa$ and large $N_{\rm T}$)
where in the last line we used the fact
that the trace of an SU($3$) matrix satisfies
$|{\rm tr}_{\rm color}U|=|\sum_{i=1}^{3}\lambda_i|\le\sum_{i=1}^{3}|\lambda_i|=3$
and ${\rm tr}_{\rm spin}P^{(4)}=2$.
Assembling various pieces, we obtain an approximate upper bound given by 
\be
|\theta|
\le
12
N_{\rm L}^3
(2\kappa)^{N_{\rm T}}
\sinh(\mu/T).
\label{eqn:upperbound}
\ee

Consider the case that we keep $\mu/T$ fixed.  Then this expression tells us that the phase becomes exponentially suppressed for larger temporal extent $N_{\rm T}$, while 
it increases in proportion to the lattice spatial volume $N_{\rm L}^3$. 
%so the sign problem gets reduced.
We also observe that the phase vanishes in the static limit $\kappa\to 0$ with spatial volume fixed.
Therefore the order of the static limit and the thermodynamic limit is subtle.

Alternatively, one may wish to consider the case that $a\mu$ 
is kept fixed.  In this case, we rewrite $\mu/T=a\mu N_{\rm T}$
in eq.~(\ref{eqn:upperbound}) and find that 
\bea
|\theta|
&\le&
12
N_{\rm L}^3
(2\kappa)^{N_{\rm T}}
\sinh(a\mu N_{\rm T})
\non\\
&\sim&
6
N_{\rm L}^3
(2\kappa)^{N_{\rm T}}
\exp(a\mu N_{\rm T})
\non\\
&=&
6
N_{\rm L}^3
\exp(\{\ln(2\kappa)+a\mu\}N_{\rm T}).
\label{eqn:upperexp}
\eea
The bound still decreases exponentially for large $N_{\rm T}$ 
if $a\mu< -\ln(2\kappa)$ is satisfied.
This bound on $a\mu$ is mild since $-\ln(2\kappa) > 1.386\cdots$ for $\kappa>0.125$, and hence covers the region 
$\mu\sim 1/a\sim 1-2$~GeV in current typical simulations.
%Then we may expect the exponential dumping in the
%region $\mu<2$GeV for the cutoff $a^{-1}=2$GeV.
Suppose that we change the temperature as a function of $N_{\rm T}$ while keeping other parameters $a$, $N_{\rm L}$, $\kappa$ and $a\mu$ fixed.
By increasing $N_{\rm T}$ the upper bound in eq.~(\ref{eqn:upperexp})
becomes exponentially suppressed and the sign problem may be avoided at low temperatures.

We have seen that the magnitude of the phase becomes smaller
as the temporal lattice size $N_{\rm T}$ increases.
We still have to examine whether the magnitude of the phase 
is under control even if the spatial volume becomes large.
To this end, let us consider the situation that all physical scales, {\it viz.}, 
\bi
\item $T=1/(aN_{\rm T})$: temperature,
\item $\mu$: chemical potential,
\item $V=L^3=(aN_{\rm L})^3$: spatial volume,
\item $m$: physical mass.
\ei
are kept fixed.  
If we change the lattice spacing by a factor $b>0$,
then we have to rescale the other parameters
to keep all the physical scales constant:
\bi
\item $a\longrightarrow a/b$,
\item $N_{\rm T}\longrightarrow bN_{\rm T}$,
\item $N_{\rm L}\longrightarrow bN_{\rm L}$,
\item $a\mu\longrightarrow a\mu/b$,
\item $\kappa\longrightarrow \kappa^\prime\approx\kappa$ ($am\longrightarrow am/b$),
\ei
where we have made use of the fact that $\kappa$ is not so sensitive to the change of the lattice spacing
compared to other parameters.
%Let us see how the upper bound of the magnitude of the phase behaves by the scaling factor $b$.
The ratio of the bound before and after the scaling is given by
\be
\frac
{{\mbox{bound of }}|\theta|_{\rm after}}
{{\mbox{bound of }}|\theta|_{\rm before}}
=
\frac
{12(bN_{\rm L})^3(2\kappa^\prime)^{bN_{\rm T}}\sinh(\mu/T)}
{12N_{\rm L}^3(2\kappa)^{N_{\rm T}}\sinh(\mu/T)}
\approx
b^3(2\kappa)^{N_{\rm T}(b-1)}.
\ee
For typical values of $2\kappa$ and $N_{\rm T}$ (say $2\kappa<1$ and $N_{\rm T}>4$),
reducing the lattice spacing also reduces the magnitude of the phase since the exponential factor $(2\kappa)^{N_{\rm T}(b-1)}$ dominates over the polynomial factor $b^3$.
It is important to check whether this behavior of the phase is observed for the exact phase.
This is under investigation with numerical simulations.% [ref,Kuramashi, Nakamura, Takeda, Uakawa]?????

%3
%=================================
\section{Numerical test in $N_f=4$ QCD}
\label{sec:numericaltest}
%=================================
%
\subsection{Simulation parameters}
\label{sec:simulation}

We employ the clover-improved Wilson quark action for four degenerate flavors and the Iwasaki gauge action.
The simulation parameters are the same as in Ref.~\cite{Li:2010qf}, namely
$\beta=1.60$, $\kappa=0.1371$ and $c_{\rm sw}=1.9655$,
which corresponds to the lattice spacing $a=0.328$ fm and $m_{\pi}=834$ MeV.
The spatial lattice size is set to $N_{\rm L}=6$ which gives $L=N_{\rm L}a\approx2$ fm,
while we employ various temporal sizes of $N_{\rm T}=4,6,8$ and $12$ which
correspond to the temperature $T=1/(a N_{\rm T})=50-150$ MeV.
The quark chemical potential is also varied in the range $0.05\le a\mu\le0.8$.

We use the conventional HMC algorithm for the phase-quenched quark determinant with the iso-spin chemical potential $\mu_{\rm u}=-\mu_{\rm d}$.
Two independent pseudo-fermions are prepared to incorporate $N_{\rm f}=4$ dynamical quarks.
We set the trajectory length to unity and vary the step size $d\tau=1/50-1/240$ depending on $a\mu$ and $N_{\rm T}$
such that the HMC acceptance rate stays around $90$ \%.
Since the fermion force tends to be large for larger chemical potential, we diminish $d\tau$ to keep the HMC acceptance rate.
For each parameter set, $5000-20000$ trajectories are accumulated.
We employ the jackknife analysis for the error estimate.
%and the bin size dependence are monitored carefully.

\subsection{Convergence check and analysis of the exact phase}
\label{sec:analysis}

\begin{figure}[p]
\begin{center}
\begin{tabular}{cccc}
\scalebox{1.0}{\includegraphics{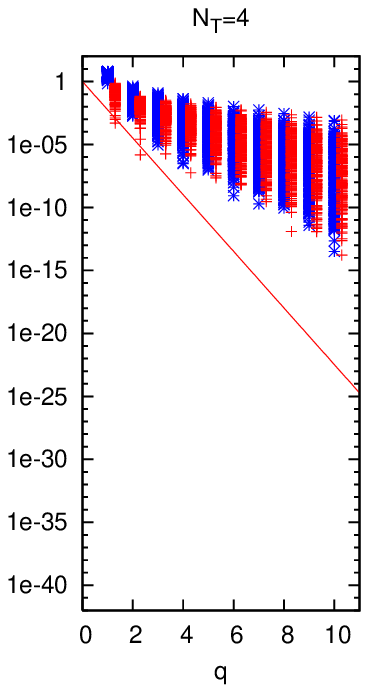}}
&
\hspace{-65mm}
\scalebox{1.0}{\includegraphics{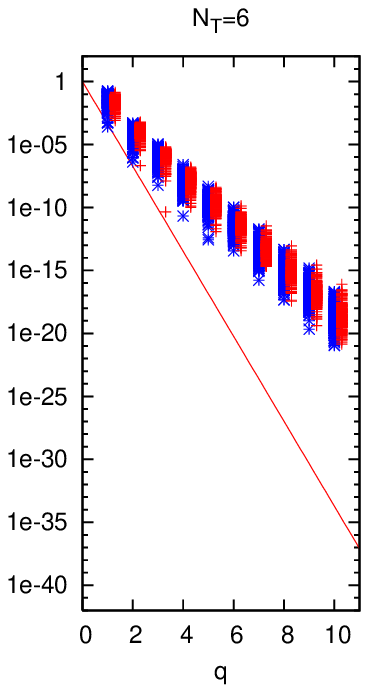}}
&
\hspace{-65mm}
\scalebox{1.0}{\includegraphics{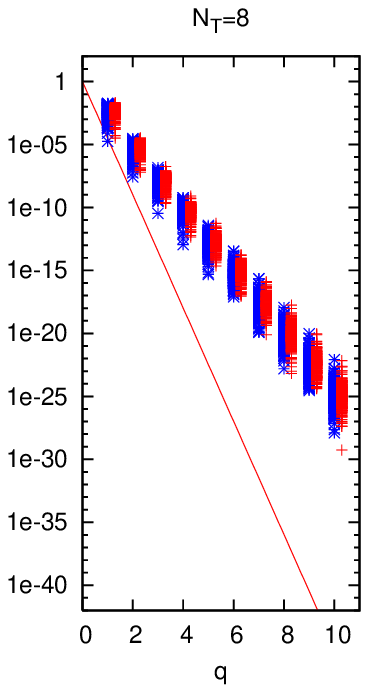}}
&
\hspace{-65mm}
\scalebox{1.0}{\includegraphics{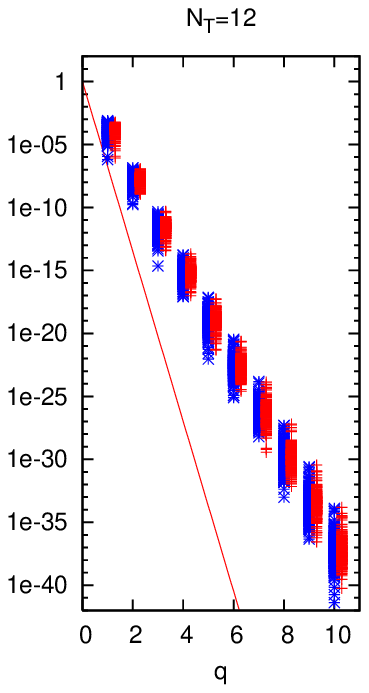}}
\\
\end{tabular}\end{center}
\caption{Absolute values of real (blue points) and imaginary (red points) parts of $V^{(q)}$
for $q=1,2,...,10$ with various $N_{\rm T}$.
They are measured at $a\mu=0.2$ employing 200 phase-quenched configurations.
Red line denotes an expected asymptotic behavior $(2\kappa)^{qN_{\rm T}}$ given in eq.~(\ref{eqn:Tqq}).
}
\label{fig:Tq}
\end{figure}

We first check the exponential decay behavior of $V^{(q)}$ as a function of the winding index $q$ expected from eq.~(\ref{eqn:Tqq}). 
Figure~\ref{fig:Tq} shows the $q$ and $N_{\rm T}$ dependence of the real and imaginary parts of $V^{(q)}$ at $a\mu=0.2$ measured on 200 phase-quenched configurations. 
We observe a clear exponential decay behavior, though the decay rate 
is somewhat milder than eq.~(\ref{eqn:Tqq}).
%Note that $V^{(q)}$ decays very quickly for larger $N_{\rm T}$.

%%%%%%%%%
%%%%%%%%%

\begin{figure}[p]
\begin{center}
\begin{tabular}{cccc}
\scalebox{1.0}{\includegraphics{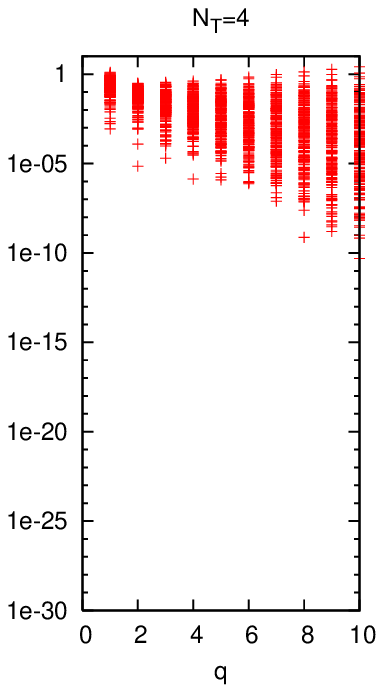}}
&
\hspace{-65mm}
\scalebox{1.0}{\includegraphics{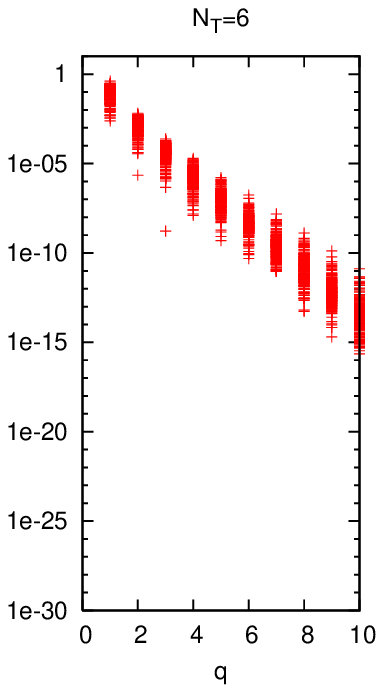}}
&
\hspace{-65mm}
\scalebox{1.0}{\includegraphics{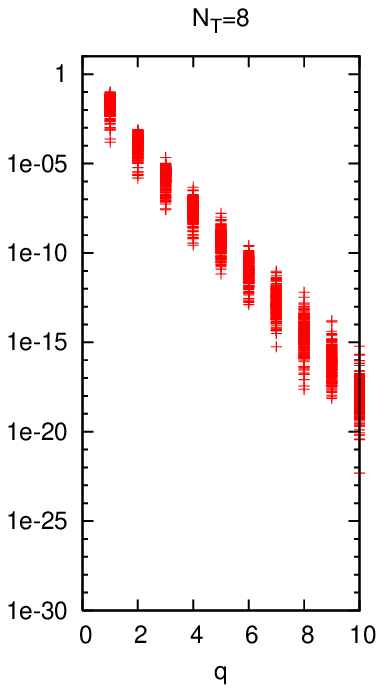}}
&
\hspace{-65mm}
\scalebox{1.0}{\includegraphics{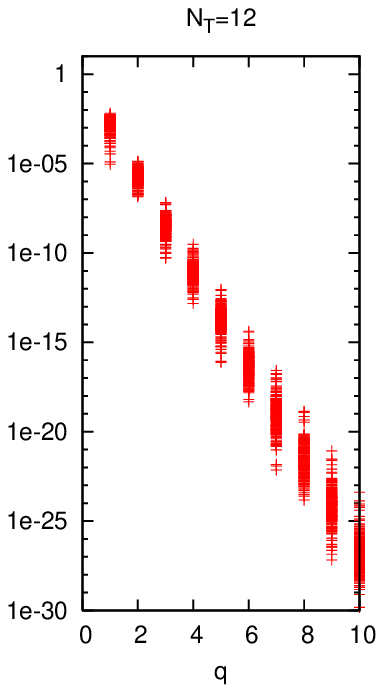}}
\\
\end{tabular}\end{center}
\caption{Absolute value of each term of the phase in eq.~(\ref{eqn:thetaTq}):
$2\sinh(q\mu/T)|{\rm Im}[V^{(q)}]|$ for $q=1,2,...,10$ with various $N_{\rm T}$
on the same configurations as those in Fig.~\ref{fig:Tq}.
}
\label{fig:phase}
\end{figure}

We next examine the behavior of the phase.  In Fig.~\ref{fig:phase} we plot the $q$ dependence of $2\sinh(q\mu/T)|{\rm Im}[V^{(q)}]|$ in eq.~(\ref{eqn:ubound_theta}).  Even after the factor $\sinh(q\mu/T)$ is combined,
the results show exponential fall-off as a function of $q$ for $N_{\rm T}\ge 6$. 
As expected, the convergence for the phase becomes better for larger $N_{\rm T}$.

%%%%%%%%%
%%%%%%%%%

\begin{figure}[p]
\begin{center}
\scalebox{1.7}{\includegraphics{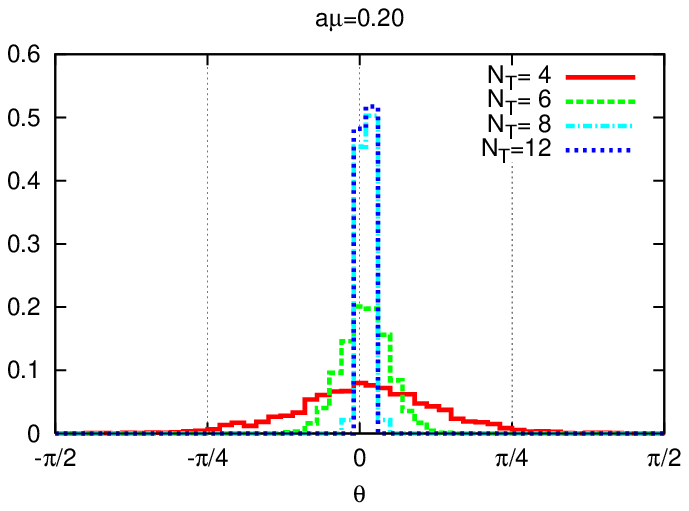}}
\end{center}
\caption{Histogram of the phase at $a\mu=0.2$ with various $N_{\rm T}$.
Other parameters are the same as those of Fig.~\ref{fig:Tq}.}
\label{fig:histogram}
\end{figure}

Finally let us examine the exact phase by calculating it directly from the quark determinant.
The calculational procedure is explained in Subsection \ref{sec:RW} together with that of quark number density.
Figure~\ref{fig:histogram} show the distribution of the phase for the single Wilson-Dirac determinant measured on the
same phase-quenched configurations as in Figs.~\ref{fig:Tq} and \ref{fig:phase}.
We find that the shape of the distribution becomes sharper for larger $N_{\rm T}$. 

\begin{figure}[p]
\begin{center}
\scalebox{1.7}{\includegraphics{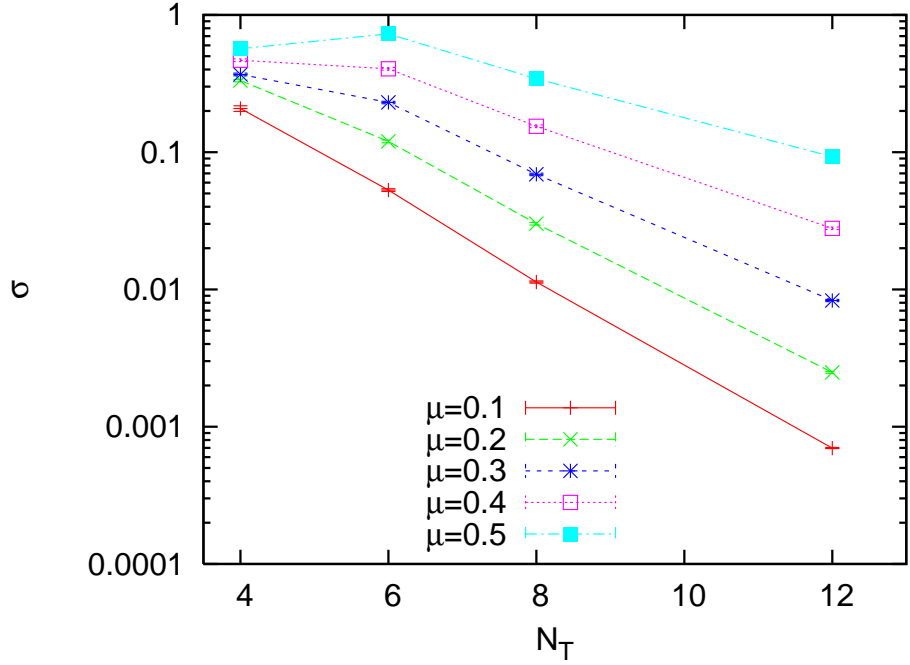}}
\end{center}
\caption{Width of the distribution of phase obtained by fitting the distribution to the normal distribution as a function of the temporal size $N_{\rm T}$ for $a\mu=0.1$ to 0.5.
Other parameters are the same as those of Fig.~\ref{fig:Tq}.}
\label{fig:width}
\end{figure}

We fit the distribution of phase by a normal distribution and extract the width $\sigma$ as a function of $a\mu$ and $N_{\rm T}$, which is plotted in Fig.~\ref{fig:width}.  The exponential decrease of the width nicely confirms our expectation. The slope of the exponential decrease is similar to that calculated from the upper bound estimate in eq.~(\ref{eqn:upperexp}).%ukawa comment: this sentence needs to be checked with data

\begin{figure}[p]
\begin{center}
\scalebox{1.7}{\includegraphics{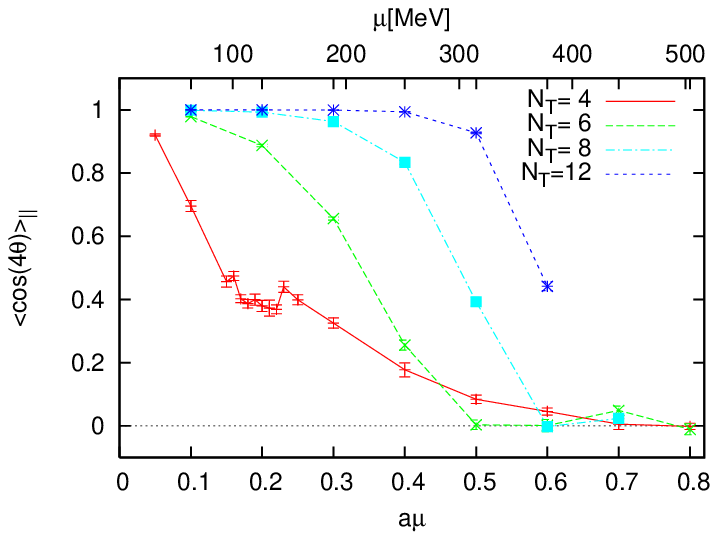}}
\end{center}
\caption{
Averaged reweighting factor as as function of $a\mu$ with various $N_{\rm T}$.
Other parameters are the same as those in Fig.~\ref{fig:Tq}.
}
\label{fig:cos}
\end{figure}

Another important quantity is the averaged phase factor $\langle e^{i4\theta}\rangle_{||}=\langle\cos(4\theta)\rangle_{||}$ where the factor $4$ accounts for the degenerate $N_{\rm f}=4$ flavors.  The average is taken on the phase-quenched configurations and the precise definition of $\langle...\rangle_{||}$ is again given in Subsection \ref{sec:RW}. Figure \ref{fig:cos} shows the result as a function of $a\mu$ for $N_{\rm T}=4,6,8, 12$ for $N_{\rm L}=6$.  
The average stays closer to unity for larger $N_{\rm T}$,
even for rather large $a\mu$.
Note that the charged pion condensation \cite{Son:2000xc} could occur
when $a\mu> a\mu_{\rm c}=am_\pi/2\approx0.7$ at $T=0$.

%%%%%%%%%
%%%%%%%%%

%%%%%%%%%

%%%%%%%%%
%%%%%%%%%

\subsection{Physical observables with reweighting method}
\label{sec:RW}

Since the phase can be controlled by increasing the temporal size of the lattice, we expect that the phase reweighting method would work in the low temperature region.
The reweighting formula for the degenerate $N_{\rm f}$-flavor case is given by
\be
\langle
{\cal O}
\rangle
=
\frac
{\langle {\cal O}{\rm e}^{iN_{\rm f}\theta}\rangle_{||}}
{\langle {\rm e}^{iN_{\rm f}\theta}\rangle_{||}},
\label{eqn:reweighting}
\ee
where the reweighted and the phase-quenched ensembles averages are defined as
\bea
\langle
{\cal O}
\rangle
&=&
\frac{\int [dU] {\rm e}^{-S_{\rm G}}(\det D)^{N_{\rm f}}{\cal O}[U]}
{\int [dU] {\rm e}^{-S_{\rm G}}(\det D)^{N_{\rm f}}},
\\
\langle
{\cal O}
\rangle_{||}
&=&
\frac{\int [dU] {\rm e}^{-S_{\rm G}}|\det D|^{N_{\rm f}}{\cal O}[U]}
{\int [dU] {\rm e}^{-S_{\rm G}}|\det D|^{N_{\rm f}}}.
\label{eqn:phasequenchedaverage}
\eea
%with the gauge link variable $U\in$ SU(3) and the lattice gauge action $S_{\rm G}[U]$.

We measure the quark number, 
the plaquette, the polyakov loop,  and their susceptibilities
by using the reweighting formula in eq.~(\ref{eqn:reweighting}).
The plaquette and the polyakov loop are measured at every trajectory,
while the quark number $n_{\rm q}$ and the susceptibility $\chi_{\rm q}$  are exactly calculated for every 10 trajectories without relying on the noise method.

The expressions for $n_{\rm q}$ and $\chi_{\rm q}$ are derived from 
the quark determinant in the reduced form given by 
\be
\det D(\mu)=A_0
\det[1-H_0-e^{\mu/T}H_+-e^{-\mu/T}H_-],
\ee
where $H_k$ with $k=0,\pm$ are given in eqs.~(\ref{eqn:H0})$-$(\ref{eqn:H-}). 
%and they are rank-$12N_{\rm L}^3$ dense matrix.
Differentiating the grand canonical partition function with respect to $\mu/T$, we derive 
\bea
\frac{n_{\rm q}}{T^3}
&=&
\frac{N_{\rm f}}{VT^3}
\langle
\tr[K^{-1}\dot{K}]
\rangle,
\\
\frac{\chi_{\rm q}}{T^2}
&=&
\frac{N_{\rm f}}{VT^3}
\left\langle
\tr[K^{-1}\ddot{K}
-
K^{-1}\dot{K}K^{-1}\dot{K}
]
\right\rangle
\non\\
&+&
\frac{N_{\rm f}^2}{VT^3}
\left\langle
\left(\tr[K^{-1}\dot{K}]\right)^2
\right\rangle
-
\frac{N_{\rm f}^2}{VT^3}
\left\langle
\tr[K^{-1}\dot{K}]
\right\rangle^2,
\eea
with
\bea
K(\mu/T)
&=&
1-H_0-e^{\mu/T}H_+-e^{-\mu/T}H_-,
\\
\dot{K}(\mu/T)&=&-[e^{\mu/T}H_+-e^{-\mu/T}H_-],
\\
\ddot{K}(\mu/T)&=&-[e^{\mu/T}H_++e^{-\mu/T}H_-]. 
\eea
The phase is given by 
\be
\tan\theta
=
\frac{{\rm Im}[\det K]}{{\rm Re}[\det K]}.
\label{eqn:tantheta}
\ee

Let us make a comment on the computational cost.  The matrices $H_\pm$, $H_0$ are dense matrices of rank $12N_{\rm L}^3$.  Once they are calculated and stored in memory, the phase
and the quark number and its susceptibility can be calculated exactly up to machine precision.
The computational cost of $H_k$, in turn, scales as $(N_{\rm L}^3(N_{\rm T}/2-1))^3$ due to the LU decomposition
for the inversion of the matrices $D_{(1)}$ and $D_{(3)}$ in eq.~(\ref{eqn:Dwb}) which have rank $12N_{\rm L}^3(N_{\rm T}/2-1)$. 
It is crucial to reduce these inversion costs, and we have implemented a further reduction, achieving a computational cost proportional to $N_{\rm L}^9N_{\rm T}$.  The required memory scales as $N_{\rm L}^6$.

%%%%%%%%%
\begin{figure}[p]
\begin{center}
\begin{tabular}{c}
\scalebox{1.7}{\includegraphics{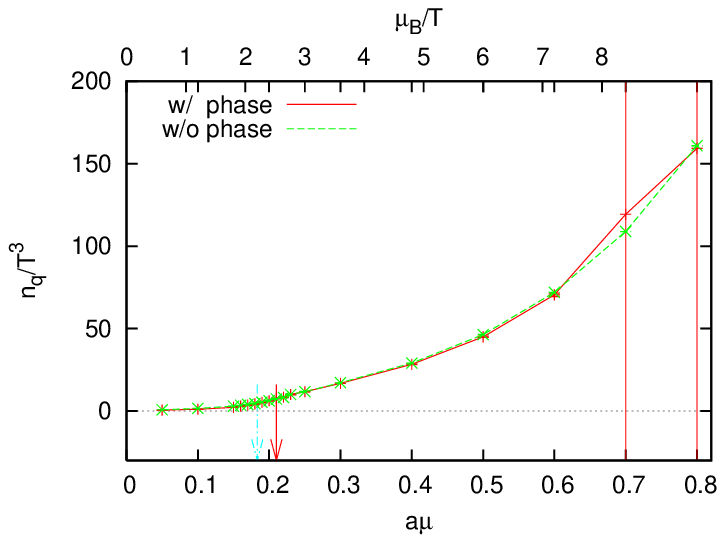}}
\vspace{10mm}
\\
\scalebox{1.7}{\includegraphics{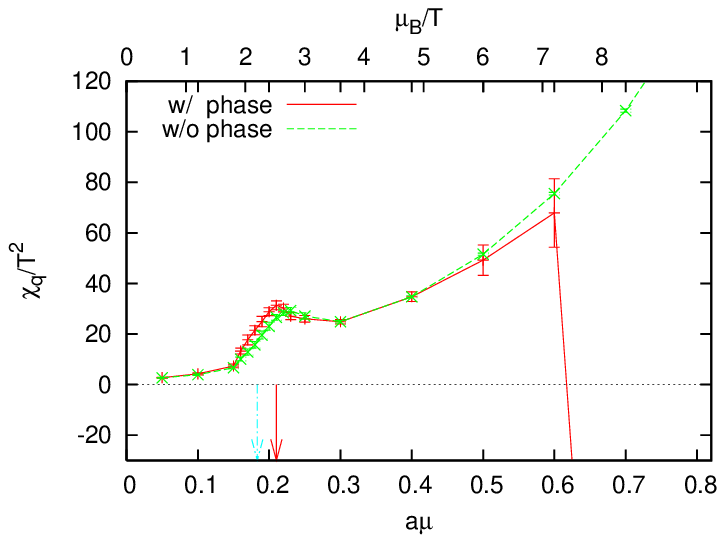}}
\end{tabular}
\end{center}
\caption{
Quark number density (upper) and
its susceptibility (lower) as a function of $a\mu$ (or the baryon chemical potential $\mu_{\rm B}=3\mu$
in unit of the temperature for upper horizontal axis)
with and without the phase factor.
%SBG (Stefan Boltzman gas) denotes high temperature limit.
%HRG (hadron resonance gas) is for low temperature limit.
%For $n_{\rm q}/T^3$ of HRG, $F(T)$ in eq.~(\ref{eqn:nHRG}) is determined to be 0.75 by a fit in the range $0.05\le a\mu\le0.15$ (3 points).
%For $\chi_{\rm q}/T^2$ of HRG in eq.(\ref{eqn:chiHRG}), we use $F(T)$ determined above as input.
Light blue dotted arrow shows the location of transition point
determined by the canonical approach \cite{Li:2010qf}.
Red solid arrow shows the location of a peak
of the quark number susceptibility ($a\mu=0.21$).
The resluts at $a\mu=0.7$ and $0.8$ have the huge error bar
because of the extremely small reweighting factor (an appearance of the sign problem)
as shown in Fig.~\ref{fig:cos}.
}
\label{fig:QN4}
\end{figure}
%%%%%%%%%

%%%%%%%%%
\begin{figure}[p]
\begin{center}
\begin{tabular}{c}
\scalebox{1.7}{\includegraphics{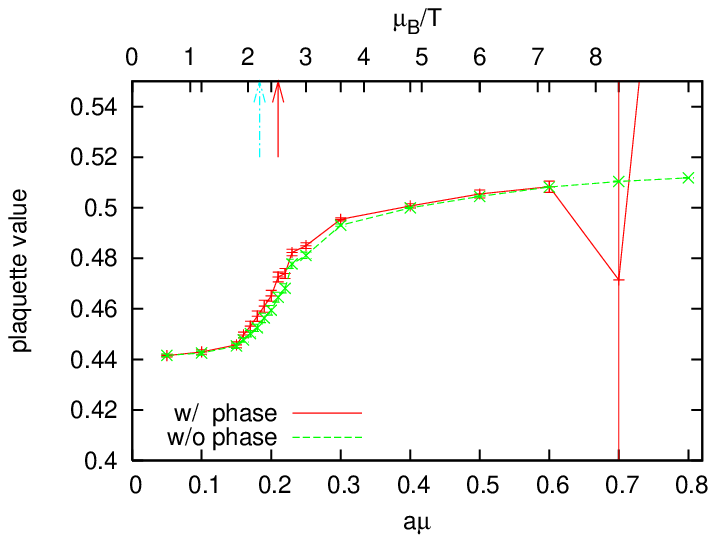}}
\vspace{10mm}
\\
\scalebox{1.7}{\includegraphics{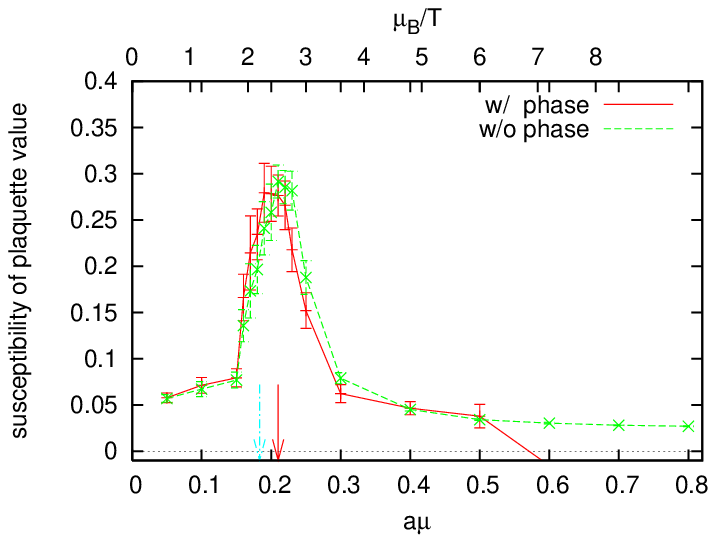}}
\end{tabular}
\end{center}
\caption{
Averaged plaquette value (upper) and its susceptibility (lower)
as a function of $a\mu$
with and without phase factor, namely quark and iso-spin chemical potential cases.
}
\label{fig:plaqmu4}
\end{figure}
%%%%%%%%%

%%%%%%%%%
\begin{figure}[p]
\begin{center}
\begin{tabular}{c}
\scalebox{1.7}{\includegraphics{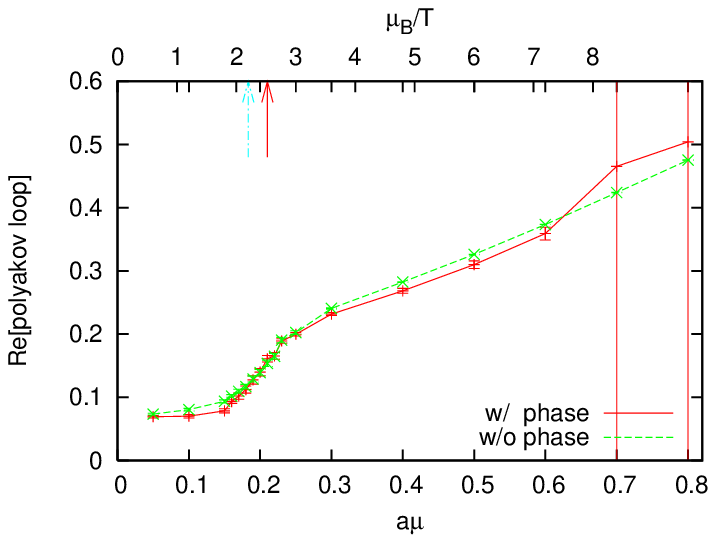}}
\vspace{10mm}
\\
\scalebox{1.7}{\includegraphics{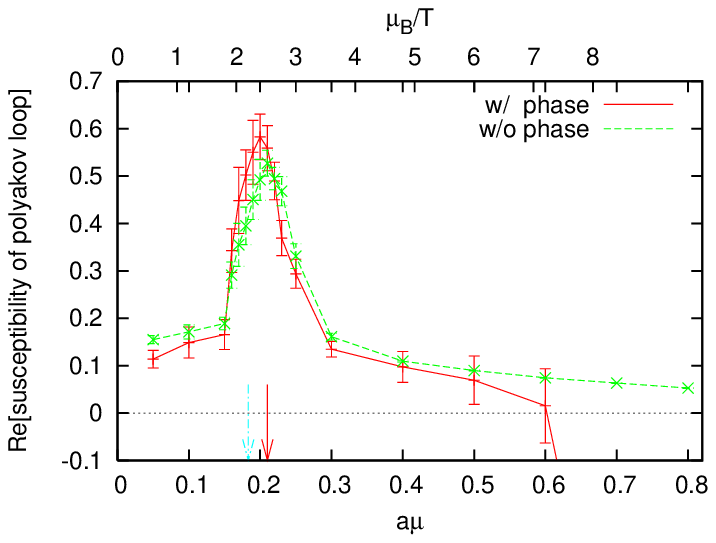}}
\end{tabular}
\end{center}
\caption{
Averaged real part of the polyakov loop (upper) and its susceptibility (lower) as a function of $a\mu$
with and without the phase factor.
}
\label{fig:polymu4}
\end{figure}
%%%%%%%%%

%%%%%%%%%
\begin{figure}[p]
\begin{center}
\begin{tabular}{ccc}
\scalebox{1.0}{\includegraphics{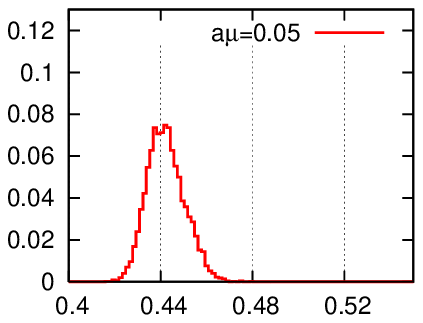}}
&
\scalebox{1.0}{\includegraphics{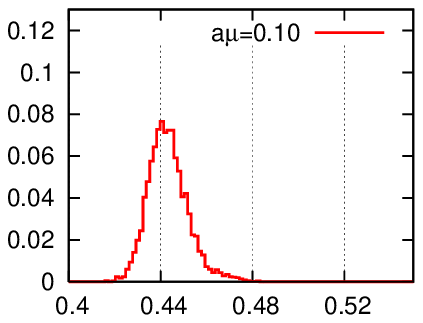}}
&
\scalebox{1.0}{\includegraphics{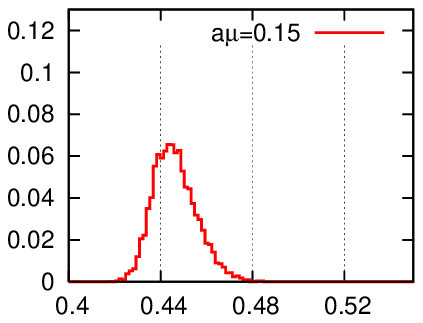}}
\\
\scalebox{1.0}{\includegraphics{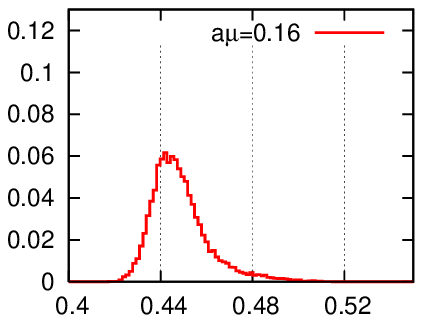}}
&
\scalebox{1.0}{\includegraphics{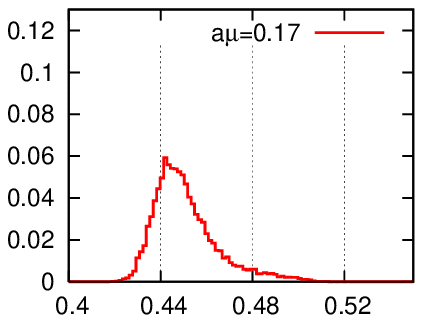}}
&
\scalebox{1.0}{\includegraphics{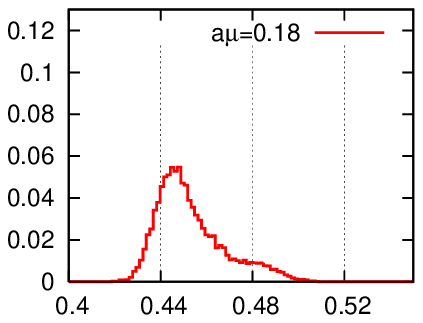}}
\\
\scalebox{1.0}{\includegraphics{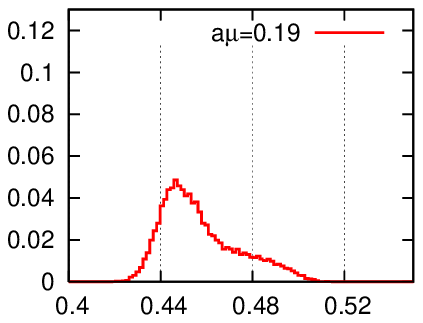}}
&
\scalebox{1.0}{\includegraphics{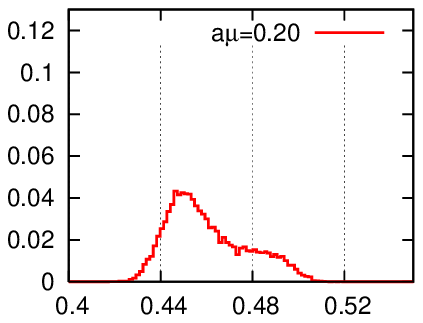}}
&
\scalebox{1.0}{\includegraphics{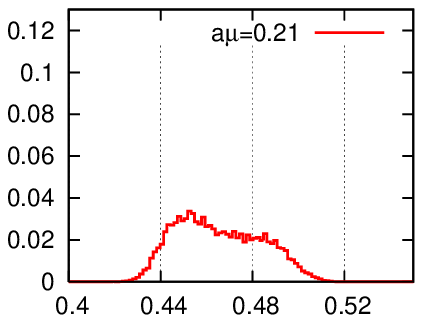}}
\\
\scalebox{1.0}{\includegraphics{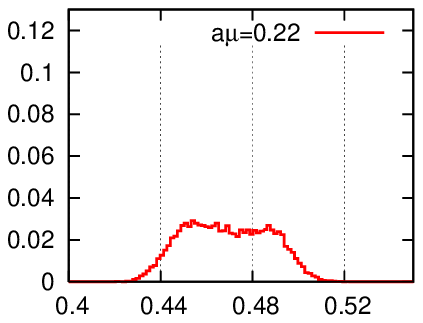}}
&
\scalebox{1.0}{\includegraphics{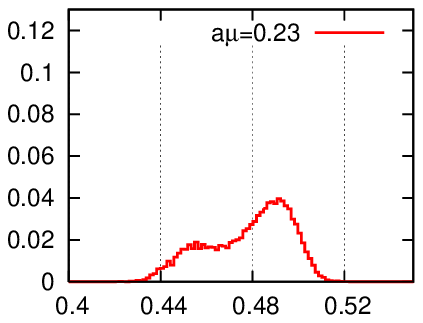}}
&
\scalebox{1.0}{\includegraphics{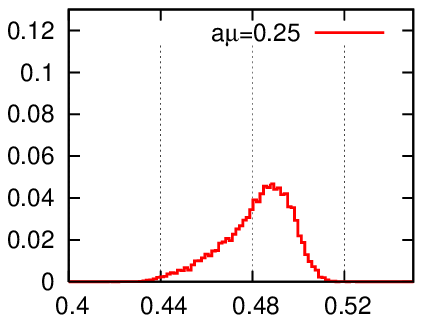}}
\\
\scalebox{1.0}{\includegraphics{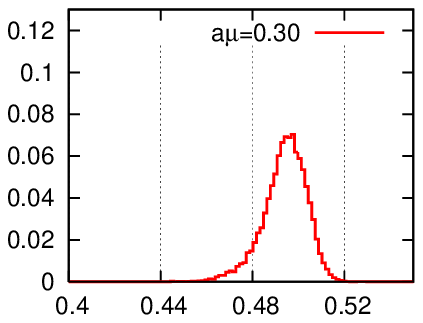}}
&
\scalebox{1.0}{\includegraphics{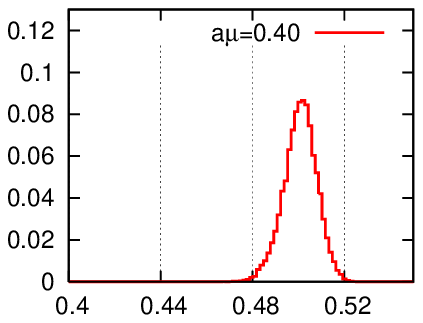}}
&
\scalebox{1.0}{\includegraphics{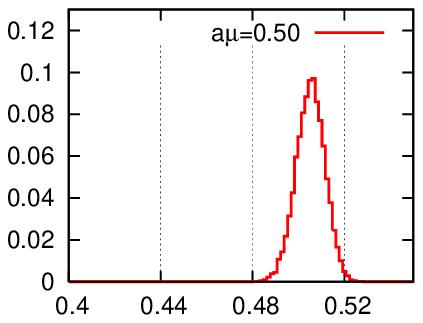}}
\\
%\scalebox{0.6}{\includegraphics{histogram_plaq_NL6_NT4_MU0.60.eps}}
%&
%\scalebox{0.6}{\includegraphics{histogram_plaq_NL6_NT4_MU0.70.eps}}
%&
%\scalebox{0.6}{\includegraphics{histogram_plaq_NL6_NT4_MU0.80.eps}}
%\\
\end{tabular}
\end{center}
\caption{
Histogram of plaquette value on the phase-quenched configuration
for $N_{\rm T}=4$ with various $a\mu$.
}
\label{fig:plaqhistogram}
\end{figure}
Figure~\ref{fig:QN4} shows our results for the quark number and its susceptibility for $N_{\rm T}=4$.  Reweighted averages are shown by red symbols and connected by red solid line ("with phase"), whereas the phase quenched averages are shown in green symbols and connected by green dashed line ("without phase").
%  The analytic results in eqs.~(\ref{eqn:nSBG})$-$(\ref{eqn:chiHRG}) are also drawn, by blue dotted lines for free quark gas and by red dotted lines for hadron resonance gas (we only included nucleon with experemental mass).%ukawa comment: which baryon did you include to compute HRG curve?  
We observe that the effect of the phase is not very large for $a\mu$ up to 0.6 over which the average of the phase stays non-zero (see Fig.~\ref{fig:cos}).  
The peak of the susceptibilities slightly shift to lower $a\mu$ for the reweighted average.  
A similar phenomenon is observed also for the plaquette and the polyakov loop 
shown in Figs.~\ref{fig:plaqmu4} and \ref{fig:polymu4},  respectively.
At $a\mu=0.7$ and beyond, the normal average loses control since the phase average becomes consistent with zero.  
%This is somehow consistent with the claim of ????[ref] that the phase quenched theory
%and full theory are similar.

The presence of the peak observed in the susceptibilities suggests a phase transition.  
Numerically the peaks are located at $a\mu\approx 0.21$ where $n_{\rm q}/T^3\approx 7.7$. In terms of the variables $\mu_{\rm B}/T$,  where $\mu_{\rm B}=3\mu$ is the baryon chemical potential, and the baryon number $n_{\rm B}$ used in Fig.~7 ($T=0.95T_{\rm c}$) of Ref.~\cite{Li:2010qf}, these values translate into $\mu_{\rm B}/T\approx 2.5$ and $n_{\rm B}\approx 8.6$, which are roughly consistent with the region of S shape found by the canonical approach in Ref.~\cite{Li:2010qf}.  
%We believe that their tiny difference will disappear in the thermodynamic limit.(ukawa comment: I do not understand this statement. Which tiny difference is discussed here?)

%For the quark number density and susceptibility, our results and the hadron resonance gas show an agreement at small $\mu$ below the peak region of the susceptibilities.  On the other hand, in the region of higher $\mu$ beyond the peak, the free massless quark curve is roughly 4--5 times smaller than the measured values.  This may due to heavy quark mass and a large lattice spacing used in the simulation.  

The histograms of the plaquette  on the phase-quenched configuration
are shown in Fig.~\ref{fig:plaqhistogram}.
There is a double peak around $a\mu=0.21$ indicating a first order phase transition.
In order to pin down the order of transition,
one has to carry out a finite size scaling.  This is left for future work. 
%, but we do not do it here and we consider it is a future work.

We have also carried out the reweighting simulations increasing the temporal size from  $N_{\rm T}=4$ to 6, 8 and $12$,
thus lowering the temperature from $T=150$ MeV to 100MeV down to 50 MeV.
We could calculate observables reliably over a significant range of $a\mu$ since the phase average stays non-vanishing as shown in Fig.~\ref{fig:cos}.  
The results, however, showed only smooth variations, and we did not observe signals of the phase transition over the region of $a\mu$ where the phase is under control. 
It may well be that, at lower temperatures, the expected phase transition takes place at larger $a\mu$ where one runs into the sign problem, {\it i.e.}, the phase average becomes very small or even vanishes.  
This possibility reminds us of Ref.~\cite{Han:2008xj}
in which the authors showed for the two-flavor random matrix model that the phase average vanishes at the boundary of a region surrounding the phase transition line in the $(T, \mu)$ plane.

%4
%=============================================
%\section{Future direction} 
%\label{sec:future}
%=============================================
%
%***{\it Do we include this section?}***
%Plan B:
%\bi
%\item |Re(det)| is taken into the boltzman factor
%\item uggly force, next slover
%\item reference Hsu and Forcrand.
%\item avoid overlap problem but not sign problem
%\ei

%5
%=============================================
\section{Concluding remarks} 
\label{sec:conclusion}
%=============================================
%
We  studied the phase of the quark determinant in QCD for  finite chemical potential.
Analytical estimates show that the winding number expansion
has a nice convergence property, and we have confirmed this through a numerical calculation of the expansion terms in phase-quenched QCD.  We hope that this expansion proves to be useful to explore the high density region at low temperatures, which has not been explored well by numerical simulations. 
%On the basis of this line,
%we propose a method that the absolute value of the real part of the determinant
%is incorporated in the molecular dynamics force of HMC.
%This may remove the overlap problem, although there is still the sign problem.

We also found that
the magnitude of the phase becomes smaller for larger temporal size of the lattice. We have shown this both by analytical estimates based on the hopping parameter expansion and by an exact evaluation of the determinat. 
The quark mass which we employ in this paper is quite heavy, however.  Hence we have to
check whether this property persists toward the physical quark mass region.
As a further result, 
simple scaling argument combined with the hopping parameter expansion of the phase leads to the prediction that 
the phase becomes smaller for smaller lattice spacing
with all the physical scales fixed.
This is an interesting possibility which we hope to check in future.

Finally, we explored the phase structure for $N_{\rm f}=4$ QCD 
with the use of 
the reweighting method, and found signals indicative of a first order phase transition on an $6^3\times 4$ lattice at  $T\approx 150$ MeV and $\mu\approx 110$ MeV.  This result is consistent with that of Ref.~\cite{Li:2010qf} using canonical methods. 
Needless to say, in order to pin down the order of the transition,
one has to carry out the finite size scaling.
%We consider this is a future work.

%As a final comment,
%after completing the remaining works given above
%we will proceed to more physically relevant situation $N_{\rm f}=3$.

%
%=====================
\section*{Acknowledgments}
%=====================
%
The authors gratefully acknowledges the useful conversation
with Mike Endress, Yoshifumi Nakamura, Sinya Aoki, Kazuyuki Kanaya and Shinji Ejiri.
We thank Ken-Ichi Ishikawa for providing us his code used in this work.
This work is supported in part by the Grants-in-Aid for
Scientific Research from the Ministry of Education, 
Culture, Sports, Science and Technology 
(Nos.
23105707, %takeda shingakujutsu
23740177, %takeda wakate B
22244018, %kura kiban A
20105002). %kura shingakujyutsu
% A part of this research has been funded by MEXT HPCI STRATEGIC PROGRAM.
The numerical calculations have been done on T2K-Tsukuba and T2K-Tokyo cluster system
at University of Tsukuba and University of Tokyo respectively.
%
%=======
\appendix
%=======
%
%=====================
\section{Reduction for the quark determinant of the Wilson-Dirac operator}
\label{sec:wilsonoperator}
%=====================
%
To fix and summarize our notations,
we briefly review a reduction technique in Ref.~\cite{Danzer:2008xs}.

\subsection{Wilson-Dirac operator}

%%%%%%%%%
\begin{figure}[p]
\begin{center}
\psfragscanon
\psfrag{L4}[c][c][2][0]{$\Lambda_{(4)}$}
\psfrag{L3}[c][c][2][0]{$\Lambda_{(3)}$}
\psfrag{L2}[c][c][2][0]{$\Lambda_{(2)}$}
\psfrag{L1}[c][c][2][0]{$\Lambda_{(1)}$}
\psfrag{N_T}[l][l][1.2][0]{$N_{\rm T}$}
\psfrag{N_T-1}[l][l][1.2][0]{$N_{\rm T}-1$}
\psfrag{N_T/2+1}[l][l][1.2][0]{$N_{\rm T}/2+1$}
\psfrag{N_T/2}[l][l][1.2][0]{$N_{\rm T}/2$}
\psfrag{N_T/2-1}[l][l][1.2][0]{$N_{\rm T}/2-1$}
\psfrag{1}[l][l][1.2][0]{$1$}
\psfrag{time}[l][l][1.5][0]{time}
\psfrag{space}[l][l][1.5][0]{space}
\scalebox{0.7}{\includegraphics{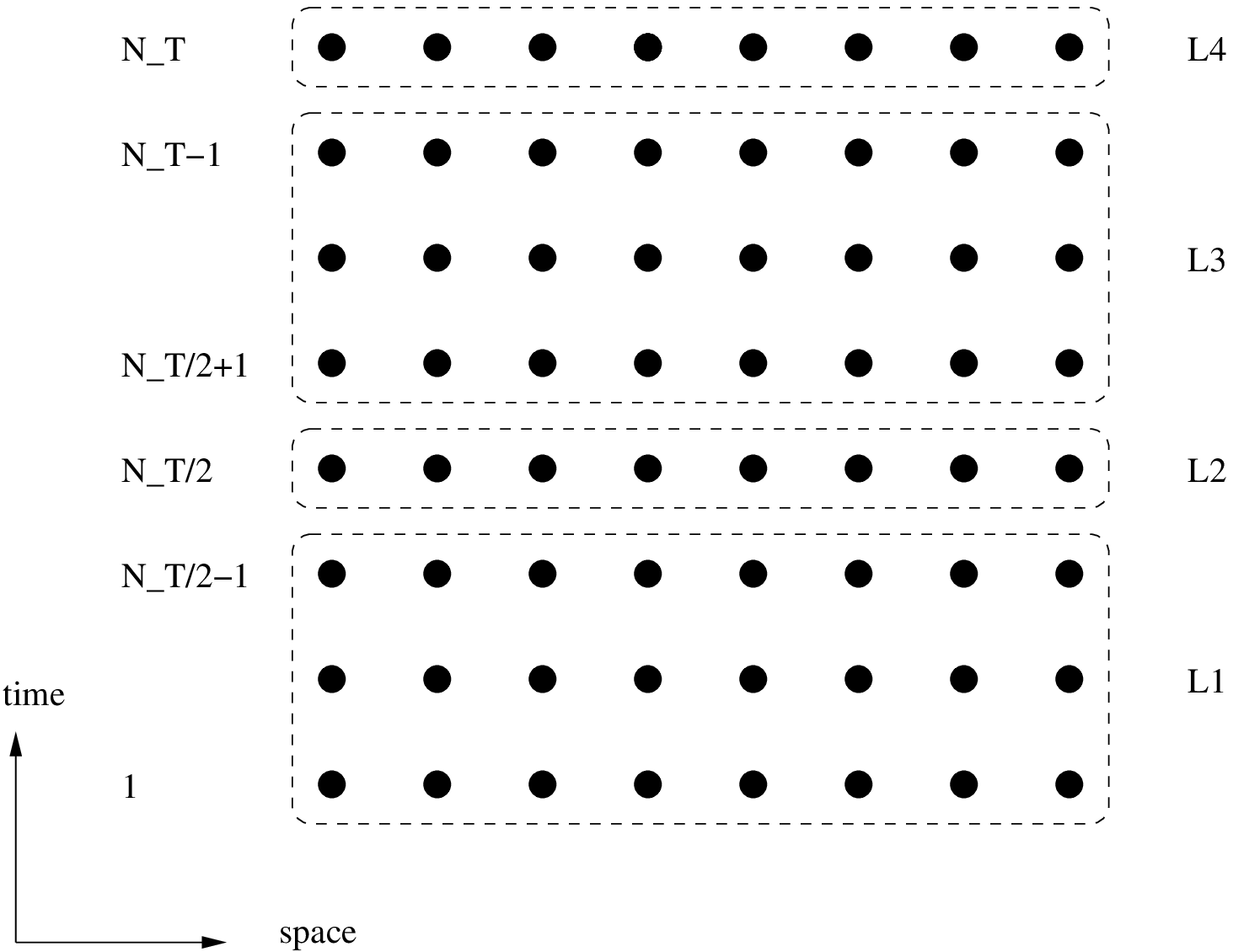}}
\caption{Domain decomposition in two dimensional case.
Time index $x_4$ runs from $1$ to $N_{\rm T}$.
\label{fig:domains}
}
\end{center}
\end{figure}
%%%%%%%%%

We consider the Wilson-Dirac operator\footnote{Inclusion of
the clover term is straightforward.}
on the lattice
with the size $N_{\rm L}^3\times N_{\rm T}$.
The Wilson fermion action is given by
\bea
S_{\rm W}
&=&
\sum_x
\left(
\bar\psi(x)\psi(x)
\right.
\non
\\
&&
-2\kappa
\sum_{k=1}^3
[
 \bar\psi(x)P_-^{(k)}U(x,k)\psi(x+\hat{k})
+\bar\psi(x+\hat{k})P_+^{(k)}U(x,k)^\dag\psi(x)
]
\non
\\
&&
\left.
-2\kappa
[
 e^{a\mu}\bar\psi(x)P_-^{(4)}U(x,4)\psi(x+\hat{4})
+e^{-a\mu}\bar\psi(x+\hat{4})P_+^{(4)}U(x,4)^\dag\psi(x)
]
\right),
\eea
with the projections $P_\pm^{(\nu)}=(1\pm\gamma_\nu)/2$.
The fermion fields satisfy the anti-periodic boundary conditions
\bea
\psi({\bf x},N_{\rm T}+1)
&=&
-\psi({\bf x},1),
\\
\bar\psi({\bf x},N_{\rm T}+1)
&=&
-\bar\psi({\bf x},1).
\eea
Applying the follwing transformation to the fermion fields,
\bea
\psi({\bf x},x_4)
&\longrightarrow&
e^{-a\mu x_4}
\psi({\bf x},x_4),
\\
\bar\psi({\bf x},x_4)
&\longrightarrow&
e^{a\mu x_4}
\bar\psi({\bf x},x_4),
\eea
the $\mu$-dependence is localized in the time-hopping
term at $x_4=1$ and $x_4=N_{\rm T}$ planes.

By decomposing the above lattice
in the time direction as illustrated in Fig.~\ref{fig:domains},
\bea
\Lambda_{(1)}&:& \mbox{for } x_4=1,2,3,..,N_{\rm T}/2-1,
\\
\Lambda_{(2)}&:& \mbox{for } x_4=N_{\rm T}/2,
\\
\Lambda_{(3)}&:& \mbox{for } x_4=N_{\rm T}/2+1,N_{\rm T}/2+2,...,N_{\rm T}-1,
\\
\Lambda_{(4)}&:& \mbox{for } x_4=N_{\rm T},
\eea
the decomposed operator can be written as (time blocked form)
\be
D(\mu)
=
\left[
\begin{array}{c|c|c|c}
D_{(1)}&D_{(12)}&0&e^{-\mu/T}D_{(14)}\\
\hline
D_{(21)}&D_{(2)}&D_{(23)}&0\\
\hline
0&D_{(32)}&D_{(3)}&D_{(34)}\\
\hline
e^{\mu/T}D_{(41)}&0&D_{(43)}&D_{(4)}\\
\end{array}
\right],
\label{eqn:Dwb}
\ee
where the $\mu$-dependence is shown explicitly.
Note that $D_{(i)}$ and $D_{(ij)}$ do not depend on $\mu$
and follow the $\gamma_5$-hermiticity and $\gamma_5$-relation
\bea
{D_{(i)}}^{\dag}
&=&
\gamma_5 D_{(i)}\gamma_5,
\hspace{5mm}\mbox{ for } i=1,2,3,4,
\label{eqn:Digamma5}
\\
{D_{(ij)}}^{\dag}
&=&
\gamma_5 D_{(ji)}\gamma_5,
\hspace{5mm}\mbox{ for } (ij)=(12),(23),(34),(41).
\label{eqn:Dijgamma5}
\eea
This decomposition is useful
for the factorization of fermion determinant
as we will see later.

\subsection{Reduction for the time direction}
By making use of the following formula repeatedly
\bea
\det
\left[
\begin{array}{cc}
A&B\\
C&D\\
\end{array}
\right]
&=&
\det [A] \det [D] \det [1-D^{-1}CA^{-1}B],
\eea
one can obtain the reduced expression of the determinant,
\bea
\det D(\mu)
&=&
\det D_{(1)}
\det D_{(3)}
\det D_{(2*2)}
\det D_{(4*4)}
\non\\
&\times&
\det
[
1-D_{(4*4)}^{-1}(e^{\mu/T}D_{(412)}+D_{(432)})D_{(2*2)}^{-1}(e^{-\mu/T}D_{(214)}+D_{(234)})
],
\label{eqn:detreduction}
\eea
where we have defined
\bea
D_{(2*2)}&=&D_{(2)}-D_{(21)}D_{(1)}^{-1}D_{(12)}-D_{(23)}D_{(3)}^{-1}D_{(32)},
\\
D_{(4*4)}&=&D_{(4)}-D_{(41)}D_{(1)}^{-1}D_{(14)}-D_{(43)}D_{(3)}^{-1}D_{(34)},
\\
D_{(412)}&=&D_{(41)}D_{(1)}^{-1}D_{(12)},
\\
D_{(432)}&=&D_{(43)}D_{(3)}^{-1}D_{(32)},
\\
D_{(214)}&=&D_{(21)}D_{(1)}^{-1}D_{(14)},
\\
D_{(234)}&=&D_{(23)}D_{(3)}^{-1}D_{(34)}.
\eea
From eqs.~(\ref{eqn:Digamma5}) and (\ref{eqn:Dijgamma5})
they have the $\gamma_5$-hermiticity and the $\gamma_5$-relation,
\be
{D}_{(i*i)}^{\dag}
=
\gamma_5 D_{(i*i)}\gamma_5  \mbox{ for } i=2,4,
\label{eqn:Diigamma5}
\ee
\be
D_{(ikj)}^{\dag}
=
\gamma_5 D_{(jki)}\gamma_5  \mbox{ for } i,j=2,4 \mbox{ and } k=1,3.
\label{eqn:Dikjgamma5}
\ee

\begin{figure}[p]
\begin{center}
\psfragscanon
\psfrag{L4}[c][c][2][0]{$\Lambda_{(4)}$}
\psfrag{L3}[c][c][2][0]{$\Lambda_{(3)}$}
\psfrag{L2}[c][c][2][0]{$\Lambda_{(2)}$}
\psfrag{L1}[c][c][2][0]{$\Lambda_{(1)}$}
\psfrag{H0}[c][c][2][0]{$H_0$}
\psfrag{H+}[c][c][2][0]{$H_+$}
\psfrag{H-}[c][c][2][0]{$H_-$}
\psfrag{time}[l][l][1.5][0]{time}
\psfrag{space}[l][l][1.5][0]{space}
\scalebox{0.7}{\includegraphics{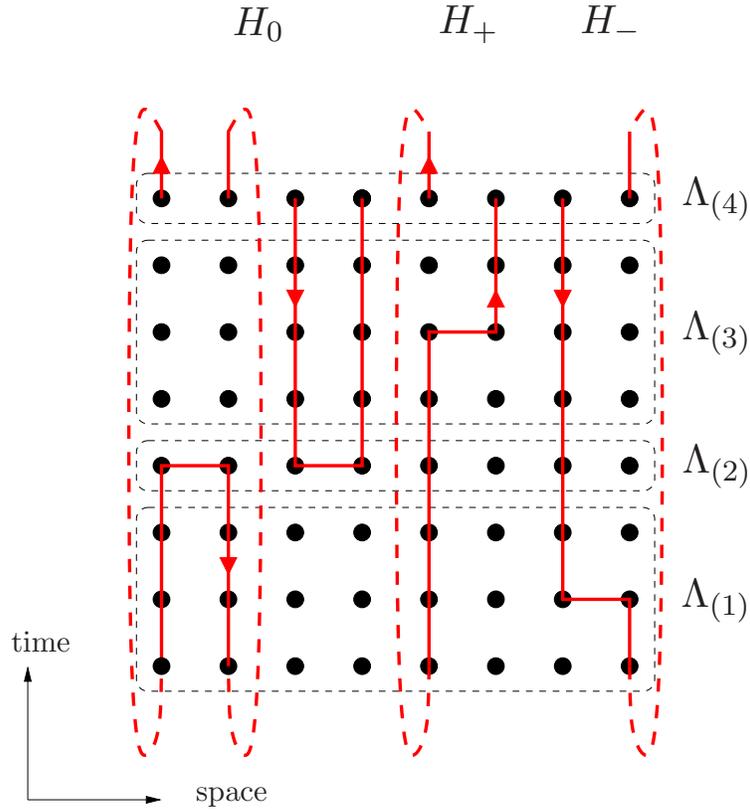}}
\caption{Geometrical pictures of $H_0$, $H_\pm$.
\label{fig:Hk}
}
\end{center}
\end{figure}

The expression of eq.~(\ref{eqn:detreduction}) is simplified as
\be
\det D(\mu)
=
A_0
\det
[1-H_0-e^{\mu/T}H_+-e^{-\mu/T}H_-],
\label{eqn:WHHH}
\ee
with the introduction of
\bea
A_0
&=&
\det D_{(1)} \det D_{(3)} \det D_{(2*2)} \det D_{(4*4)},
\label{eqn:A0}
\\
H_0
&=&
 D_{(4*4)}^{-1}D_{(412)}D_{(2*2)}^{-1}D_{(214)}
+D_{(4*4)}^{-1}D_{(432)}D_{(2*2)}^{-1}D_{(234)},
\label{eqn:H0}
\\
H_+
&=&
 D_{(4*4)}^{-1}D_{(412)}
 D_{(2*2)}^{-1}D_{(234)},
\label{eqn:H+}
\\
H_-
&=&
 D_{(4*4)}^{-1}D_{(432)}
 D_{(2*2)}^{-1}D_{(214)}.
\label{eqn:H-}
\eea
From eqs.~(\ref{eqn:Digamma5}) and (\ref{eqn:Diigamma5}),
$A_0$ is shown to be real.
An geometrical meaning of $H_k$ ($k=0,\pm$) is as follows (see also Fig.~\ref{fig:Hk}):
\bi
\item $H_0$: paths through domains
$\Lambda_{(4)}\rightarrow \Lambda_{(1)} ({\rm or} \Lambda_{(3)})\rightarrow \Lambda_{(2)} \rightarrow \Lambda_{(1)}({\rm or} \Lambda_{(3)}) \rightarrow \Lambda_{(4)}$\\
$\Longrightarrow$ NO winding around time direction.
\item $H_+$: paths through domains
$\Lambda_{(4)}\rightarrow \Lambda_{(1)} \rightarrow \Lambda_{(2)} \rightarrow \Lambda_{(3)} \rightarrow \Lambda_{(4)}$\\
$\Longrightarrow$ Forward winding around time direction.
\item $H_-$: paths through domains
$\Lambda_{(4)}\rightarrow \Lambda_{(3)} \rightarrow \Lambda_{(2)} \rightarrow \Lambda_{(1)} \rightarrow \Lambda_{(4)}$\\
$\Longrightarrow$ Backward winding around time direction.
\ei
From eqs.~(\ref{eqn:Diigamma5}) and (\ref{eqn:Dikjgamma5})
we find that they have $\gamma_5$-relation
\be
H_k^{\dag}
=
\gamma_5 D_{(4*4)}H_{-k}D_{(4*4)}^{-1}\gamma_5
\label{eqn:Hkgamma5}
\ee
for $k=0,\pm$.

%%%%%%%%%
%%%%%%%%%

\subsection{Winding number expansion}

We can rewrite $\det D(\mu)$ in eq.~(\ref{eqn:WHHH}) as follows,
\bea
\det D(\mu)&=&
A_0
\det [1-H_0-e^{\mu/T}H_+-e^{-\mu/T}H_-]
\\
&=&
A_0
\exp\left(
{\rm Tr}\ln[1-H_0-e^{\mu/T}H_+-e^{-\mu/T}H_-]
\right)
\\
&=&
A_0
\exp\left(
-\sum_{n=1}^{\infty}\frac{1}{n}{\rm Tr}[H_0+e^{\mu/T}H_++e^{-\mu/T}H_-]^n
\right)\\
&=&
A_0
\exp\left(
-\sum_{q\in \mathbb{Z}}^{\infty}e^{q\mu/T}\sum_{n=1}^{\infty}\frac{1}{n}\sum_{k_1+...+k_n=q}{\rm Tr}[H_{k_1}H_{k_2}...H_{k_n}]
\right),
\eea
where $k_i\in\{0,\pm 1\}$ and
$q$ is an integer which counts how many times the individual loop
${\rm Tr}[H_{k_1}H_{k_2}...H_{k_n}]$ winds around the time direction.
In order to simplify the equation,
we introduce
\bea
V^{(q)}
&=&
\sum_{n=1}^{\infty}V_n^{(q)},
\label{eqn:Tq}
\\
V_n^{(q)}
&=&
\frac{1}{n}\sum_{k_1+...+k_n=q}{\rm Tr} [H_{k_1}H_{k_2}...H_{k_n}],
\label{eqn:Tqn}
\eea
where $k_i\in\{0,\pm 1\}$.
Note that $V_n^{(q)}=0$ for $q>n$ by definition.
Since for an actual evaluation of $V^{(q)}$
one cannot of course sum up all the terms in eq.(\ref{eqn:Tq}),
we adopt the truncation scheme in Ref.~\cite{Danzer:2008xs},
\bea
\hat{V}^{(0)}
&=&
{\rm tr}[H_0]
+
\frac{1}{2}
{\rm tr}[(H_0)^2]
+
{\rm tr}[H_+H_-],
\\
\hat{V}^{(q)}
&=&
\frac{1}{q}
{\rm tr}[(H_+)^q]
+
{\rm tr}[(H_+)^qH_0],
\hspace{8mm}
\mbox{ for }
q=1,2,3,...
\eea
The relative truncation errors are $O((2\kappa)^{N_{\rm T}})$
thus can be safely neglected.
From eq.(\ref{eqn:Hkgamma5})
one find
\be
{\rm Tr} [H_{k_1}H_{k_2}...H_{k_n}]
=
{\rm Tr} [H_{-k_n}...H_{-k_2}H_{-k_1}]^{\ast}.
\label{eqn:H}
\ee
This yields
\bea
V_n^{(0)}&\in&\mathbb{R},
\\
V_n^{(-q)}&=&V_n^{(q)\ast}.
\label{eqn:Tnast}
\eea
Thus one can show
eqs.~(\ref{eqn:Tqast0}) and (\ref{eqn:Tqast}).
Finally we summarize the winding number expansion of the logarithm
of the determinant
\bea
\det D(\mu)
&=&
A_0
\exp
\left[
-\sum_{q\in\mathbb{Z}}
e^{q\mu/T}V^{(q)}
\right].
\eea

%\newpage
%
%%%%%%%%%%%%% Bibliography
%\bibliographystyle{physics_tst}
%\bibliography{biblio}

\end{document}